\documentclass[twocolumn,showpacs,preprintnumbers,amsmath,amssymb]{revtex4}
\usepackage{epsfig}
\usepackage{amsmath}
\usepackage{epsfig}
\usepackage{psfrag}
\usepackage{amsfonts}
\usepackage{graphicx}% Include figure files
\usepackage{dcolumn}% Align table columns on decimal point
\usepackage{bm}% bold math
\usepackage{lineno}% line numbers

\usepackage{graphicx}% Include figure files
\usepackage{dcolumn}% Align table columns on decimal point
\usepackage{bm}% bold math
\newcommand{\vect}{\left ( \begin{array}{c}}
\newcommand{\evect}{\end{array} \right )}

\usepackage{epsfig}
\usepackage{psfrag}
\usepackage{graphicx}% Include figure files
\usepackage{dcolumn}% Align table columns on decimal point
\usepackage{bm}% bold math
\def\fsl#1{\setbox0=\hbox{$#1$}                 % set a box for #1
   \dimen0=\wd0                                 % and get its size
   \setbox1=\hbox{/} \dimen1=\wd1               % get size of /
   \ifdim\dimen0>\dimen1                        % #1 is bigger
      \rlap{\hbox to \dimen0{\hfil/\hfil}}      % so center / in box
      #1                                        % and print #1
   \else                                        % / is bigger
      \rlap{\hbox to \dimen1{\hfil$#1$\hfil}}   % so center #1
      /                                         % and print /
   \fi}                                         %

\begin{document}
\title{A Possible Higher Order Correction to the Chiral Vortical Conductivity in a Gauge Field Plasma}

\author{De-fu Hou}
\affiliation{Institute of Particle Physics,Key Laboratory of QLP of  MOE, Huazhong Normal
University, Wuhan 430079, China}
\author { Hui Liu }
\affiliation{ Physics Department, Jinan University, Guangzhou, China }
\author{Hai-cang Ren}\affiliation{Physics Department, The Rockefeller University,
1230 York Avenue, New York, NY 10021-6399}
\affiliation{Institute of Particle Physics,Key Laboratory of QLP of  MOE, Huazhong Normal University, Wuhan 430079, China}

\begin{abstract}

The two loop contributions to the chiral vortical conductivity are considered. The Kubo formula together
with the anomalous Ward identity of the axial vector current suggest that there may be a nonzero
correction to the coefficient of the $T^2$ term of the conductivity.

\end{abstract}
%\preprint{ {\normalsize RU09-5-A} }

\pacs{12.38.Aw, 24.85.+p, 26.60.+c}

%\keywords{Chiral vortical effect,  higher order correction}
\maketitle

The chiral magnetic effect and the chiral vortical effect have been actively investigated for recent years.
Because of the triangle anomaly, an external magnetic field and/or an fluid vorticity will induce an electric
current, a baryon current and an axial vector current in a relativistic plasma. These currents will lead to separations
of electric charges, the baryon numbers and chirality, which may be observed in the quark-gluon plasma created
through heavy ion collisions \cite{Fukushima,Khareev}. To the order of the linear response, we have
\begin{eqnarray}
\vec J_{\rm em} &=& \sigma_{\rm em}^B\vec B+\sigma_{\rm em}^V\vec\omega\nonumber\\
\vec J_{\rm b} &=& \sigma_{\rm b}^B\vec B+\sigma_{\rm b}^V\vec\omega\nonumber\\
\vec J_5 &=& \sigma_5^B\vec B+\sigma_5^V\vec\omega.
\end{eqnarray}
for the currents driven by the magnetic field and the fluid vorticity.
The anomalous transport coefficients $\sigma$'s above have been exploreded from field theoretic
point of view and from the holographic method \cite{Fukushima, Yee, Erdmenger, Banerjee, Torabian, Son, Kirsch,Lin, Nair}.
An important question along the former approach is if these coefficients are free from the higher order
corrections of coupling constants, like their origin, the triangle anomaly. In case of the Chiral Magnetic Effect(CME),
the nonrenormalization of $\sigma_{\rm em}^B$ in the homogeneous limit of a static magnetic field
has been established \cite{Rubakov,Hong, Hou} and the classical expression \cite{Fukushima}
\begin{equation}
\sigma_{\rm em}^B=N_c\sum_fq_f^2\frac{e^2\mu_5}{2\pi^2}
\end{equation}
holds to all orders of electromagnetic and $SU(N_c)$ gauge coupling,
where $N_c$ is the number of colors, $q_f$ is the charge number of each flavor and $\mu_5$ is the chemical potential of
the axial charge. The same conclusion for CME can also be reached following the argument in \cite{Golkar}.
In this note, we shall address
the parallel issue for the chiral vortical conductivity $\sigma_5^V$ to see whether it is subject to higher order
corrections.

The anomalous transport coefficient $\sigma_5^V$ was first introduced in \cite{Son} where the anomalous Ward identity
together with the 2nd law of thermodynamics yields for a relativistic plasma with
an axial charge chemical potential $\mu_5$ yields the expression
\cite{fn}

\begin{equation}
\sigma_5^V=\frac{\mu_5^2}{2\pi^2}
\end{equation}
It is realized soon after in \cite{Neiman} that the general solution to the thermodynamic condition employed
in \cite{Son} is given by
\begin{equation}
\sigma_5^V=\frac{\mu_5^2}{2\pi^2}+cT^2
\label{general}
\end{equation}
with $c$ a undetermined constant. Then came the Kubo formula \cite{Landsteiner1}
and the one-loop calculation in \cite{Landsteiner2} which confirms
the general structure (\ref{general}) and yields $c=\frac{1}{12}$. This result is also confirmed by kinetic theories
\cite{Gao}. The authors of \cite{Landsteiner2, Landsteiner3, Neiman2} related the $T^2$ term to the gravity anomaly
and a recent analysis \cite{Jensen} from a geometric point of view within a general hydrodynamical framework
suggests the nonrenormalization of the $T^2$ term. But a field theoretic aspect
regarding the higher corrections remains murky.

In a recent work \cite{Golkar}, the authors addressed the issue based on diagrammatic analysis. They generalized the
Coleman-Hill theorem \cite{Coleman} to the stress tensor insertion and proved the nonrenormalization of $\sigma_5^V$ for a $\sigma$
model. As to gauge theories, they argued that the nonrenormalization remains valid in the large $N_c$ limit because
of the structure of the anomaly. Upon a close examination of their argument for a gauge theory plasma at the
two-loop level, we found a diagram that is not covered. We shall point out this diagram and compute its contribution
to $\sigma_5^V$ below.

For the sake of clarity, we shall consider a QED plasma with the Lagrangian density
\begin{equation}
{\cal L}=-\frac{1}{4e_0^2}V^{\mu\nu}V_{\mu\nu}-i\bar\psi\gamma^\mu D_\mu\psi+\frac{1}{2}h^{\mu\nu}T_{\mu\nu}
+A^\mu J_{5\mu},
\label{qed}
\end{equation}
where $V_{\mu\nu}=\partial_\mu V_\nu-\partial_\nu V_\mu$ is the electromagnetic field tensor with $V_\mu$ in gauge
potential, the covariant derivative
\begin{equation}
D_\mu=\partial_\mu-iV_\mu
\end{equation}
and we have added couplings to an external axial vector field $A^\mu$ and a metric perturbation $h^{\mu\nu}$.
with the axial vector current
\begin{equation}
J_{5\mu}=i\bar\psi\gamma_\mu\gamma_5\psi
\end{equation}
and the stress tensor
\begin{eqnarray}
T_{\mu\nu} &=& V_\mu^{\rho} V_{\nu\rho}-\frac{1}{4}\eta_{\mu\nu}V^{\rho\lambda}V_{\rho\lambda}+
\frac{1}{4}(-D_\mu\bar\psi\gamma_\nu\psi-D_\nu\bar\psi\gamma_\mu\psi)\nonumber\\
&+&\frac{1}{4}( \bar\psi\gamma_\mu D_\nu\psi+\bar\psi\gamma_\nu D_\mu\psi).
\end{eqnarray}
We have set $A_\mu=h_{\mu\nu}=0$ in the expression of $T_{\mu\nu}$ above. The anomalous Ward identity of $J_{5\mu}$
reads
\begin{equation}
\partial_\mu J_5^\mu=\frac{e_0^2}{16\pi^2\sqrt{-g}}\epsilon^{\mu\nu\rho\lambda}V_{\mu\nu}V_{\rho\lambda}
\label{anomaly}
\end{equation}
with $g$ the determinant of the metric $g_{\mu\nu}=\eta_{\mu\nu}+h_{\mu\nu}$.

Following \cite{Landsteiner1}, the chiral vortical conductivity $\sigma_5^V$ is given by the correlators between the axial
current density and energy flux density as ${\cal G}_{ij}(Q)=\sigma_5^V\epsilon_{ijk}q_k$  in the limit
$Q=(0,\vec q)\to 0$, where
\begin{equation}
{\cal G}_{ij}(Q)=-\int_0^\infty dt \int d\vec r e^{-i\vec q\cdot\vec r}
\frac{{\rm Tr}\{e^{-\beta H}[J_{5i}(\vec r,t),T_{0j}(0,0)]\}}{{\rm Tr}e^{-\beta H}}
\label{kubo}
\end{equation}
and can be evaluated perturbatively in terms of thermal diagrams, where
$H$ the Hamiltonian corresponding to the Lagrangian density (\ref{qed}) at $A_\mu=h_{\mu\nu}=0$. All two-loop
diagrams are shown in Fig.1. The one-particle reducible diagram (g) does not contribute since the loop attached
to the axial vector vertex vanishes, as can be checked explictly. We have the two-loop contribution to
${\cal G}_{ij}(Q)$
\begin{equation}\label{green}
{\cal G}_{ij}^{(2)}(Q)={\cal G}_{ij}^{(a)}(Q)+{\cal G}_{ij}^{(b-f)}(Q)
\end{equation}
with
\begin{equation}
{\cal G}_{ij}^{(a)}(Q)=T\int\frac{d^3\vec p}{(2\pi)^2}\sum_{p_0}\Lambda_{i\alpha\beta}(P,Q)
D_{\alpha,0\gamma}\left(P_-\right)D_{j\gamma,\beta}\left( P_+\right)
\label{green1}
%\label{btof}
\end{equation}
with $P_-=P-\frac{Q}{2}, P_+=P+\frac{Q}{2}$, and
\begin{equation}
{\cal G}_{ij}^{(b-f)}(Q)=T\int\frac{d^3\vec p}{(2\pi)^2}\sum_{p_0}\Lambda_{ij\alpha\beta}(P,Q)
D_{\alpha\beta}(P)
\label{green2}
\end{equation}
where $p_0=2n\pi T$ with $n=0,\pm1, \pm2, ...$
is the Matsubara energy and the photon propagators $D_{\mu\nu}(P)$, $D_{\rho,\mu\nu}(P)$ and $D_{\mu\nu,\rho}(P)$
are given by
\begin{eqnarray}
D_{\mu\nu}(P) &=& \frac{1}{P^2}\Big[\delta_{\mu\nu}+(\kappa-1)\frac{P_\mu P_\nu}{P^2}\Big]\nonumber\\
D_{\rho,\mu\nu}(P) &=& -\frac{1}{P^2}(P_\mu\delta_{\rho\nu}-P_\nu\delta_{\rho\mu})\nonumber\\
D_{\mu\nu,\rho}(P) &=& D_{\rho,\mu\nu}(-P)
\end{eqnarray}
with $P^2=\vec p^2+p_0^2$ and $\kappa$ the gauge parameter. The momenta in (\ref{green}) are all Euclidean with
the metric $\delta_{\mu\nu}$. The amplitudes $\Lambda_{i\alpha\beta}(P,Q)$ and $\Lambda_{ij\alpha\beta}(P,Q)$
of (\ref{green1}) and (\ref{green2}) are related to the anomalous triangle diagram
$\Pi_{\mu\alpha\beta}(K_1,K_2)$, and the kernel of its metric perturbation
$\Pi_{\mu\alpha\beta,\rho\lambda}(Q,K_1,K_2)$, depicted in Fig.2, via
\begin{eqnarray}
\Lambda_{i\alpha\beta}(P,Q) &=& \Pi_{i\alpha\beta}\left(P+\frac{Q}{2},-P+\frac{Q}{2}\right)\nonumber\\
\Lambda_{ij\alpha\beta}(P,Q) &=& \Pi_{i\alpha\beta,0j}(Q,P,-P).
\end{eqnarray}
If there were no axial anomaly, the sum of all diagrams (a)-(f) would be of the order $O(q^2)$
in the limit $Q=(0,\vec q)\to 0$ according to the Coleman-Hill like argument employed in
\cite{Golkar}. As to the contribution from the
anomaly, following an elegant argument of \cite{Golkar}, sum of that from Fig. 2(b-f) couples only to
the trace of the metric perturbation. Therefore the anomaly does not contribute the diagrams Fig.1(b)-(f) with the
insertion of an off-diagonal component. The anomaly contribution to diagram Fig.1(a), however, is not covered by
the above argument and has to be examined separately.

The anomalous Ward identity (\ref{anomaly}) implies that
\begin{equation}
q_i\Lambda_{i\alpha\beta}(P,Q)=-\frac{e_0^2}{2\pi^2}\epsilon_{\alpha\beta\rho i}P^\rho q_i
\end{equation}

Taking the derivative with respect to momentum $\vec q$ on both sides, we derive
\begin{equation}
\Lambda_{i\alpha\beta}(P,Q)=-\frac{e_0^2}{2\pi^2}\epsilon_{\alpha\beta\rho i}P_\rho
-q_j\frac{\partial}{\partial q_i}\Lambda_{j\alpha\beta}(P,Q)
\end{equation}
In the absence of infrared divergence, we end up with a nonzero limit as $\vec q\to 0$,
\begin{equation}
\Lambda_{i\alpha\beta}(P,Q)\to-\frac{e_0^2}{2\pi^2}\epsilon_{\alpha\beta\rho i}P_\rho
\label{inhomo}
\end{equation}

Inserting this nonzero limit into (\ref{green1}), we find the anomaly contribution
\begin{equation}
{\cal G}_{ij}^{\rm anom}(Q)=-\frac{e_0^2T}{2\pi^2}\epsilon_{\alpha\beta\nu i}
\int\frac{d^3\vec p}{(2\pi)^2}\sum_{p_0}P_\nu
D_{\alpha,0\gamma}\left(P_-\right)D_{j\gamma,\beta}\left(P_+\right).
\label{Danom}
\end{equation}

Dropping the terms beyond linear order in $\vec q$, we obtain that
\begin{eqnarray}
{\cal G}_{ij}^{\rm anom}(Q)&= & \frac {e_0^2T}{2\pi^2}\int\frac{d^3\vec p}{(2\pi)^2}\sum_{p_0}\frac{1}{(\vec p^2+p_0^2)^2}\nonumber \\
\cdot [-\frac{1}{2}\epsilon_{ikl}p_l q_k p_j&+&(\vec p^2+p_0^2)\epsilon_{ijk}p_k-\frac{1}{2}p_0^2\epsilon_{ijk}q_k]\nonumber \\
&=& \sigma_5^{V(2)}\epsilon_{ijk}q_k
\end{eqnarray}
with $\sigma_5^{V(2)}$ the two-loop contribution to CVE coefficient given by
\begin{equation}
\sigma_5^{V(2)}=\frac{e_0^2T}{4\pi^2}\sum_{p_0}\int\frac{d^3\vec p}{(2\pi)^3}
\frac{\frac{1}{3}\vec p^2-p_0^2}{(\vec p^2+p_0^2)^2}
\label{cve}
\end{equation}

In the last step, we have dropped the 2nd term in the numerator of the integrand because it is odd in $P$, and
have replaced $p_lp_j$  by $\frac{1}{3}\vec p^2\delta_{lj}$. The integral of (\ref{cve}) can be
calculated by dimensional regularization. We have $\sigma_5^{V(2)}=\lim_{d\to 3}\sigma_{5,d}^{V(2)}$ with
\begin{eqnarray}
\sigma_{5,d}^{V(2)} &=& \frac{e_0^2T}{4\pi^2}\sum_{p_0}\int\frac{d^d\vec p}{(2\pi)^d}\frac{\frac{1}{d}p^2-p_0^2}{(p^2+p_0^2)^2} \nonumber \\
&=& \frac{e_0^2T}{16\pi^2}\frac{(d-1)\omega_d}{2^d\pi^{d-1}\sin\frac{\pi d}{2}}\sum_{p_0}|p_0|^{d-2} \nonumber \\
&=& \frac{e_0^2T^{d-1}}{32\pi^3}\frac{(d-1)\omega_d}{\sin\frac{\pi d}{2}}\zeta(2-d)
\end{eqnarray}
where $\omega_d$ is the solid angle in $d$ dimensions. Therefore
\begin{equation}
\sigma_5^{V(2)}=\frac{e_0^2}{48\pi^2}T^2
\label{higher}
\end{equation}
and the coefficient $c$ of (\ref{general}) takes the form
\begin{equation}
c=\frac{1}{12}+\frac{e_0^2}{48\pi^2}
\label{creno}
\end{equation}

Because of the universality of the axial anomaly, the second term above
are intact if the fermion number and the axial charge chemical potentials are switched on. In another word, the $\mu_5^2$
of (\ref{general}) is not renormalized by higher order terms and our result is not
in contradiction with the thermodynamic argument of \cite{Son}.

To convince ourselves the robustness of this result, we have also evaluated $\sigma_5^{V(2)}$ a la Pauli-Villars
like regularization, which amounts to $\sigma_5^{V(2)}=\lim_{M_s\to\infty}\sigma_{5,M}^{V(2)}$ with
\begin{eqnarray}
\sigma_{5,M}^{V(2)}&=&\frac{e_0^2T}{4\pi^2}\sum_{p_0}\int\frac{d^3\vec p}{(2\pi)^3}
\Big[\frac{\frac{1}{3}\vec p^2-p_0^2}{(\vec p^2+p_0^2)^2}\nonumber\\
&-&\sum_s C_s\frac{\frac{1}{3}\vec p^2-p_0^2}{(\vec p^2+p_0^2+M_s^2)^2}\Big],
\end{eqnarray}
where the coefficients $C_s$ are chosen to make the integral and the summation divergence free. On writing
\begin{eqnarray}
\sigma_{5,M}^{V(2)}&=&\frac{e_0^2}{4\pi^2}\lbrace\int\frac{d^4\vec P}{(2\pi)^4}[...]\nonumber\\
&&+\left(T\sum_{p_0}-\int_{-\infty}^\infty\frac{dp_0}{2\pi}\right)\int\frac{d^3\vec p}{(2\pi)^3}[...]\rbrace,
\end{eqnarray}

we have for the first term inside the bracket
\begin{eqnarray}
\int\frac{d^4\vec P}{(2\pi)^4}[...]&=&
\int\frac{d^4\vec P}{(2\pi)^4}\left(\frac{1}{3}\vec P^2-\frac{4}{3}p_0^2\right)\nonumber\\
\cdot\Big[\frac{1}{(\vec p^2+p_0^2)^2}
&-&\sum_s C_s\frac{1}{(\vec p^2+p_0^2+M_s^2)^2}\Big]=0
\end{eqnarray}
with $P^2=\vec p^2+p_0^2$ because of the 4d rotational symmetry once the integral is made convergence by the regulators.
As to the rest terms, following the standard
treatment of the summation over $p_0$ in terms of a contour integral, we find
\begin{eqnarray}
&&\lim_{M_s\to\infty}\left(T\sum_{p_0}-\int_{-\infty}^\infty\frac{dp_0}{2\pi}\right)
\int\frac{d^3\vec p}{(2\pi)^3}[...]\nonumber\\
&&=\int\frac{d^3\vec p}{(2\pi)^3}\Big[\frac{2}{3}\frac{e^{\frac{p}{T}}}{(e^{\frac{p}{T}}-1)^2}
-\frac{1}{3p}\frac{1}{e^{\frac{p}{T}}+1}\Big]=\frac{T^2}{12}
\end{eqnarray} 
and confirm (\ref{higher}).
\begin{figure}[h]
\centering
\includegraphics[height=1.3in]{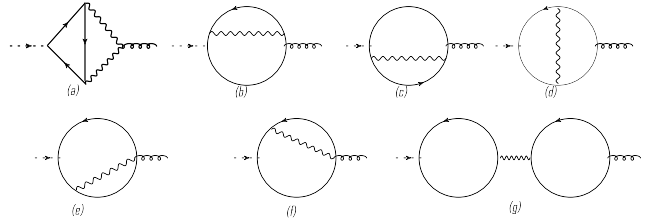}
\caption{The two-loop diagrams for the chiral vortical conductivity}
\end{figure}

\begin{figure}[h]
\centering
\includegraphics[height=1.5in]{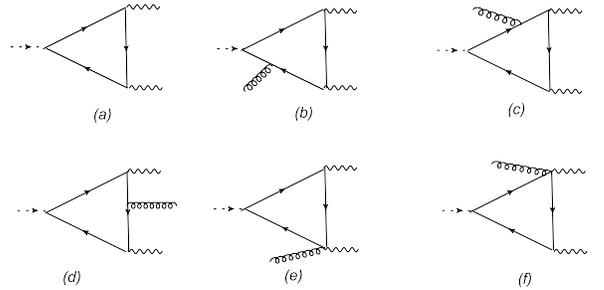}
\caption{The anomalous triangle and its metric perturbation. $\Pi_{\mu\alpha\beta}(K_1,K_2)$ in the text denotes
the amputated part of (a) and $\Pi_{\mu\alpha\beta,\rho\lambda}(Q,K_1,K_2)$ the sum of amputated parts of
(b-f), where ($\alpha$, $\beta$) refer the external photon lines with outgoing momenta ($K_1$, $K_2$), $\mu$ refers
to the axial vector insertion with incoming momentum $Q$ and ($\rho$, $\lambda$) refers to the external graviton.}
\end{figure}

Our analysis can be trivially generalized a QCD like nonAbelian gauge theory with $N_c$ colors and $N_f$ flavors.
This amounts to replace $e_0^2$ of (\ref{inhomo}) by $N_f{\rm tr}T^lT^{l'}g_0^2=\frac{1}{2}\delta^{ll'}N_fg_0^2$ with
$T^l$ the $SU(N_c)$ generator for quarks in the fundamental representation and $g_0$ the Yang-Mills coupling and to
sum (\ref{Danom}) over adjoint gluons. On writing
\begin{equation}
\sigma_5^V=N_cN_f\left(\frac{\mu_5^2}{2\pi^2}+cT^2\right),
\end{equation}
we have
\begin{equation}
c=\frac{1}{12}+\frac{N_c^2-1}{2N_c}\frac{g_0^2}{48\pi^2}
\end{equation}
and the 2nd term is not suppressed in the large $N_c$ limit for a fixed 't Hooft coupling $N_cg_0^2$. This  
makes the strong 't Hooft coupling limit nontrivial, an issue that may be addressed by the 
holographic principle.

One possible loophole with above analysis concerns the generalization of the Coleman-Hill theorem to
the stress tensor insertion. In case of the vector or axial current insertion to a diagram with only
external gauge boson lines, the transversality of the diagram post insertion can be established algebraically
prior to integrating the loop momenta \cite{Bjorken} (formally for the axial current case). We find this is not
obvious with the stress tensor insertion corresponding to one-loop diagrams with one boson line of each diagram
of Fig.1 cut open. The reason may be attributed to the fact that the external lines of the these diagrams
do not carry the conserved charges but energies and momenta. If there is no complication with the generalization
of the Coleman-Hill theorem, we do find a two loop term of the chiral vortical coefficient given by (\ref{higher}).
In any case, it would be interesting to verify
or disprove this result through an explicit calculation of all two loop diagrams of Fig.1. Alternatively, the
Matsubara formulation of the correlator in (\ref{kubo}) can also be evaluated nonperturbatively on a
lattice at $\mu_5=0$ without running into sign problems.

\begin{acknowledgments}
We are grateful to S. Golkar for email communications on their paper \cite{Golkar} and valuable comments.
We are also indebted to him for informing us that they reached a similar conclusion in their forthcoming
version of \cite{Golkar} when this work is finished.  D. Hou would like to  thank S. Lin for valuable discussions.
An email communication from K. Landsteiner is warmly acknowledged.
This work is  supported partly by NSFC under grant Nos. 10975060, 11135011, 11221504 and 10947002.
\end{acknowledgments}
%%%%%%%%%%%%%%%%%%%%%%%%%%%%%%%%%%%%%%%%

%%%%%%%%%%%%%%%%%%%%%%%%%%%%%%%%%%%%%%%%


\begin{thebibliography}{99}
\bibitem{fn} The vertical conductivity in \cite{Son} is defined in Landau frame, given by
$\frac{\mu_5^2}{2\pi^2}\left(1-\frac{2}{3}\frac{n_5\mu_5}{u+p}\right)$ with $n_5$, $u$ and $p$ the axial-charge
density, the energy density and the pressure of the fluid. The 2nd term inside the parentheses
represents the contribution from the energy flow and is subtracted out in this paper. See \cite{Landsteiner1} for
clarifications.
\end{thebibliography}
\end{document}